\documentclass[a4paper,11pt]{article}
\usepackage{bm,amsmath,amssymb,mathrsfs,graphicx,citesort}
\newcommand{\nc}{\newcommand}
\nc{\rnc}{\renewcommand}
\nc{\nn}{\nonumber}
\nc{\bra}{\langle}
\nc{\ket}{\rangle}
\rnc{\(}{\left(}
\rnc{\)}{\right)}
\rnc{\[}{\left[}
\rnc{\]}{\right]}

\usepackage[dvips]{color}
\nc{\tcr}{\textcolor{red}}

\begin{document}%
%
\title{Long time asymptotics of the totally asymmetric \\
simple exclusion process}
\author{
Kohei Motegi$^1$\thanks{E-mail: motegi@gokutan.c.u-tokyo.ac.jp} \,
Kazumitsu Sakai$^2$\thanks{E-mail: sakai@gokutan.c.u-tokyo.ac.jp} \,
and Jun Sato$^3$\thanks{E-mail: jsato@sofia.phys.ocha.ac.jp}
\\\\
$^1$Okayama Institute for Quantum Physics, \\
 Kyoyama 1-9-1, Okayama 700-0015, Japan \\
\\
$^2$Institute of physics, University of Tokyo, \\ 
Komaba 3-8-1, Meguro-ku, Tokyo 153-8902, Japan \\
\\
$^3$Department of Physics, Graduate School of Humanities and Sciences, \\
Ochanomizu University 2-1-1 Ohtsuka, Bunkyo-ku, Tokyo 112-8610, Japan
\\\\
\\
}

\date{\today}
 
 
\maketitle

%
%
\begin{abstract}
We study the long time asymptotics of the relaxation dynamics
of the totally asymmetric simple exclusion process on a ring.
Evaluating the asymptotic amplitudes of the local currents
by the algebraic Bethe ansatz method, we find the
relaxation times starting from the step and alternating initial conditions
are governed by different eigenvalues of the Markov matrix.
In both cases, the scaling exponents of the
leading asymptotic amplitudes with respect to the total number of sites
are found to be $-1$.
We also study the asymptotics of correlation functions such as the
emptiness formation probability. 
\end{abstract}


\section{Introduction}
The asymmetric simple exclusion process (ASEP)
is one of the most fundamental
models in nonequilibrium statistical mechanics \cite{De,Sch,Spit,Li,Sp}.
The ASEP is a stochastic interacting particle system
consisting of biased random walkers obeying the exclusion principle,
and have applications to biology \cite{MGP},
traffic flows \cite{Scha,SCN} and quantum dots \cite{KO}, for example.
Like the Ising model in equilibrium statistical mechanics,
the ASEP is nowadays a paradigm in nonequilibrium statistical mechanics
in the sense that many exact methods are amenable to extract various 
interesting nontrivial facts. For example, the matrix product
representation \cite{DEHP,SD,San,Sasa,BECE,USW}
of the steady state revealed boundary induced phase
transitions \cite{Kr}. For the last ten years,
due to the development of the random matrix theory 
\cite{Jo,BR,PS,NS,RS,BFPS,IS,TW},
the current fluctuations in the infinite system
were shown to satisfy the Tracy-Widom distribution.

The Bethe ansatz method,
which was originated as a traditional method to analyze 
quantum integrable models such as the Heisenberg XXZ chain,
can also be applied to analyze the ASEP.
This comes from the fact that the Markov matrix describing the
dynamics of the ASEP is equivalent to the Hamiltonian of an
integrable spin chain. Utilizing this fact,
the relaxation times and spectral gaps were examined
\cite{Dh,GS,K,GM,GM2,deGE0,deGE1,AKSS,deGFS},
and the exact expression for cumulants of currents and
large deviation functions
were obtained \cite{DL,PM,PM2,Pr,deGE2}.

One of the latest developments of the Bethe ansatz approach is 
the study of the full relaxation dynamics of the 
totally asymmetric simple exclusion process (TASEP) on a ring
by the algebraic Bethe ansatz method
\cite{MSS}.
Previously, by noting the equivalence between the
Markov matrix of the TASEP and the Hamiltonain of an integrable spin chain,
the algebraic Bethe ansatz derivation of the
Bethe ansatz equation and the construction of the determinant representation of
the scalar product \cite{TF,KBI,Bo} was done.
However, the power of the algebraic Bethe ansatz method was not fully
utilized to study the dynamics of the TASEP,
due to the difficulties of evaluating the
multipoint form factors and the overlap between the initial state
and the Bethe vector.
Formulating the dynamics of the TASEP
by evaluating the form factors and the overlap
from the algebraic Bethe ansatz,
we examined the full relaxtion dynamics of the
local densities and currents, and found
the scaling exponents of the asymptotic amplitudes 
starting from the step initial condition for example.
The Monte Carlo method can be employed to study the
full relaxation dynamics as well, but it is difficult to study the
details of the dynamics, the long time asymptotics for example.

In this paper, we study the long time behavior of the
relaxation dynamics to the steady state by the algebraic Bethe ansatz method.
We particularly focus on the local currents,
and the two fundamental initial conditions:
the step and the alternating initial conditions.
The step initial condition is the case which
the half of the system is consecutively occupied
by the particles and the other half is empty at the initial time.
The alternating initial condition is the case which
all odd sites are occupied while all even sites are empty. 
We examine the asymptotic amplitudes of the local currents,
and find that in contrast to the step initial condition,
the asymptotic amplitudes corresponding to the lowest excited
states of the Markov matrix vanish for the alternating initial condition.
In other words, the relaxation time of the alternating initial condition
is not governed by the lowest excited states.
Instead, the second lowest excited state determines the relaxation time.
Our discovery suggests that the naive guess of the relaxation time
by considering only the Markov matrix may lead to an incorrect result.

This paper is organized as follows. 
In the next section,
we review the basics of the algebraic Bethe ansatz and the scalar products.
In section 3,
we formulate the dynamics of the local densities, currents and
the emptiness formation probability of
the TASEP by the algebraic Bethe ansatz method.
This is achieved by
evaluating the form factors and the overlap between the
initial state and arbitrary Bethe vector for the case of 
step and alternating initial conditions.
The low excited states of the Markov matrix are described in section 4.
The long time asymptotics
for the step initial condition is analyzed in section 5.
In section 6, we analyze
the alternating initial condition
and extract interesting difference from the step initial condition.
Section 5 is devoted to the conclusion.

\section{Algebraic Bethe ansatz of the TASEP}
In this and the next sections, we formulate the
dynamics of the TASEP on a periodic ring by the 
algebraic Bethe ansatz method.
In this section, we review the relation between the TASEP
and the integrable spin chain, and the basics of the
algebraic Bethe ansatz and the scalar products.

\subsection{The definition of the TASEP}
We consider the TASEP on a periodic ring with $M$ sites and $N$ particles.
Since the particles obey the exclusion rule, each site
can be occupied by at most one particle.
The dynamical rule of the TASEP  is as follows:
during the time interval $\mathrm{d} t$, a particle
at a site $j$ jumps to $(j+1)$th site with probability $\mathrm{d} t$,
if the $(j+1)$th site is vacant.
For convenience, we associate a Boolean variable $\tau_i$
to every site $i$ to indicate whether a particle is 
present $(\tau_i=1)$ or not $(\tau_i=0)$.
The probability of being in the (normalized) state 
$|\tau_1, \dots, \tau_M \rangle$
is denoted as $P_t(\tau_1, \dots, \tau_M)$.
The time evolution of the state vector
$|\psi(t) \rangle=\sum_{\tau_i=0,1} P_t(\tau_1, \dots, \tau_M)
| \tau_1, \dots, \tau_M \rangle$
is subject to the master equation
\begin{align}
\frac{\mathrm{d}}{\mathrm{d} t} |\psi(t) \rangle
=\mathcal{M} |\psi(t) \rangle. \label{master}
\end{align}
Here the Markov matrix $\mathcal{M}$ of the TASEP is defined by
\begin{align}
\mathcal{M}=\sum_{j=1}^M \Bigg\{
\sigma_j^+ \sigma_{j+1}^-
+\frac{1}{4}(\sigma_j^z \sigma_{j+1}^{z}-1)
 \Bigg\}, \label{markov}
\end{align}
where $\sigma_j^\pm:=(\sigma_j^x\pm i\sigma_j^y)/2$ and
$\sigma_j^{x,y,z}$ are the Pauli matrices acting on the $j$th site.
Here we interpret the occupied $(\tau_i=1)$ and 
unoccupied $(\tau_i=0)$ state with
spin down $(|\downarrow_i \rangle)$ and up state 
$(|\uparrow_i \rangle)$, respectively.
For example, 
we interpret that $\sigma_j^z |\tau_1,\dots, \tau_M\rangle=(-1)^{\tau_j} 
|\tau_1,\dots, \tau_M\rangle$.
We denote the vacuum state (state with no particle)
$| \Omega \rangle:=|0,\dots,0 \rangle$.
Starting from any initial condition,
the state of the TASEP converges to
the steady state $| S_N \rangle$ (not normalized)
\begin{align}
| S_N \rangle=\sum_{1 \le m_1 < \cdots < m_N \le M}
\sigma_{m_1}^- \sigma_{m_2}^-
\cdots \sigma_{m_N}^-| \Omega \rangle,
\end{align}
in the long time limit.
The steady state 
is an eigenvector of the Markov matrix with zero eigenvalue
\begin{align}
\mathcal{M}|S_N \rangle=0.
\end{align}
We also define the dual vacuum state $\langle \Omega|:=\langle 0,\dots,0|$
and the left steady state vector 
\begin{align}
\langle S_N|=\sum_{1 \le m_1 < \cdots < m_N \le M}
\langle \Omega|
\sigma_{m_1}^+ \sigma_{m_2}^+
\cdots \sigma_{m_N}^+
, \label{leftsteady}
\end{align}
which is also an eigenvector of the Markov matrix
with zero eigenvalue 
\begin{align}
\langle S_N|\mathcal{M}=0.
\end{align}
The inner product between $|S_N \rangle$ and $\langle S_N|$
can be easily calculated as
\begin{align}
Z_N:=\langle S_N|S_N \rangle=
\frac{M!}{N!(M-N)!}.
\end{align}

\subsection{Algebraic Bethe ansatz}
From the spin chain point of view,
the Markov matrix (\ref{markov}), which describes the
dynamics of the TASEP,
is exactly a Hamiltonian of a one-dimensional integrable quantum spin chain.
Therefore one can use the exact methods
to examine the dynamics of the TASEP.
The Bethe ansatz  is one of the most traditional methods
to treat quantum integrable models.
The algebraic Bethe ansatz  \cite{TF,KBI,Bo}
is one of the variants of the Bethe ansatz,
which can construct the eigenvectors as well as the eigenvalues
of the Markov matrix (Hamiltonian).
Moreover, the algebraic Bethe ansatz enables us to
calculate form factors,
which are the basic ingredients of the
physical quantities.
Therefore we formulate the TASEP by the algebraic Bethe ansatz method.

\begin{figure}[tt]
\begin{center}
\includegraphics[width=0.6\textwidth]{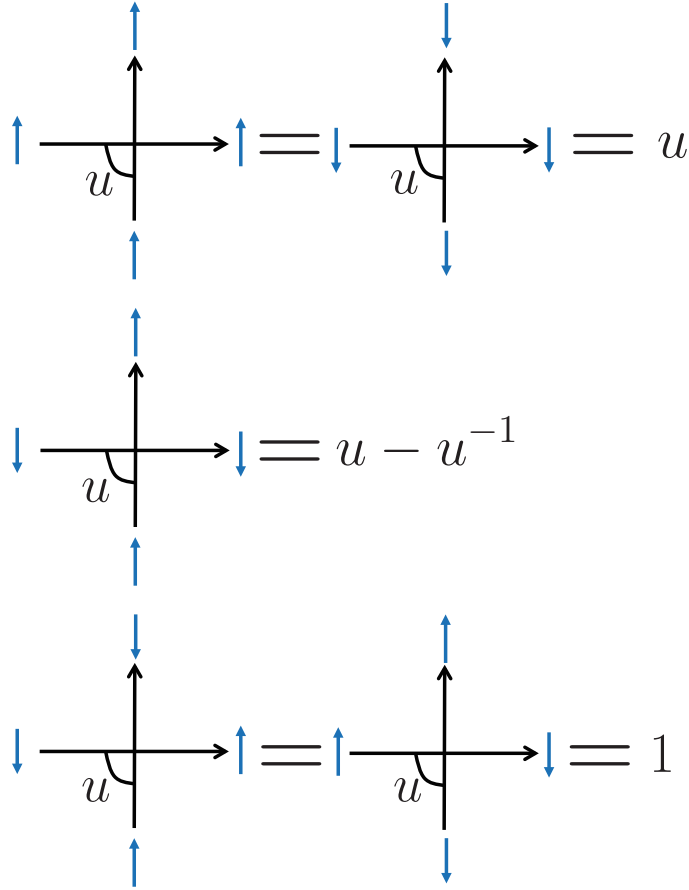}
\end{center}
\caption{The elements of the $L$-operator (\ref{loperator}).
The $L$-operator is represented as two crossing arrows.
The left and the up arrow represents an auxiliary and a quantum space
respectively.
The spins on the left and the right around a vertex
denote the input and the output of the auxiliary space,
and the ones on the bottom and the top denote the input and the
output of the quantum space, respectively. For example,
$\langle \uparrow| \langle \downarrow |_j 
L(j|u)| \downarrow \rangle | \uparrow \rangle_j=1$.
}
\label{weight}
\end{figure}

What plays a fundamental role in an integrable model is the
$L$-operator acting on the $j$th site
\begin{align}
L(j|u)
=us s_j+n(uI-u^{-1} s_j)+\sigma^- \sigma_j^+
+\sigma^+ \sigma_j^-,
\label{loperator}
\end{align}
where $s_j=(1+\sigma_j^z)/2$ and
$n_j=(1-\sigma_j^z)/2$ are the projection operator
onto the empty and filled states at $j$th site,
respectively (see figure \ref{weight} for a pictorial description).
The symbols $s$ and $n$ without subscript mean they act on the
auxiliary space.

The $L$-operator satisfies the $RLL$ relation
\begin{align}
R(u,v)(L(n|u) \otimes L(n|v))=(L(n|v) \otimes L(n|u)) R(u,v),
\label{RLL}
\end{align}
where $R(u,v)$ is the $R$-matrix
\begin{align}
R(u,v)
=&\left(
\begin{array}{cccc}
f(v,u) & 0 & 0 & 0 \\
0 & g(v,u) & 1 & 0 \\
0 & 0 & g(v,u) & 0 \\
0 & 0 & 0 & f(v,u)
\end{array}
\right), \\
f(v,u)=&\frac{u^2}{u^2-v^2}, \
g(v,u)=\frac{uv}{u^2-v^2}.
\end{align}
From the $RLL$ relation (\ref{RLL}), it follows that the
monodromy matrix which is defined as a product of
$L$ operators
\begin{align}
T(u)=\prod_{j=1}^M L(j|u)
&=\left(
\begin{array}{cc}
A(u) & B(u)  \\
C(u) & D(u)
\end{array}
\right),
\label{monodromy}
\end{align}
satisfies the $RTT$ relation
\begin{align}
R(u,v)(T(u) \otimes T(v))=(T(v) \otimes T(u))R(u,v). \label{RTT}
\end{align}
From the intertwining relation (\ref{RTT}),
one immediately finds the transfer matrix
\begin{align}
\tau(u)=u^{-M} \mathrm{tr} T(u)=u^{-M}(A(u)+D(u)),
\label{transfer}
\end{align}
forms a commuting family
\begin{align}
[\tau(u),\tau(v)]=0.
\end{align}
The Markov matrix (\ref{markov}) is 
constructed from the transfer matrix (\ref{transfer})
as
\begin{align}
\mathcal{M}=\frac{1}{2} \tau^{-1}(1) \frac{\partial}{\partial u}
\tau(u)|_{u=1}.
\end{align}
The elements of the $RTT$ relation (\ref{RTT})
give the commutation relations between the
elements of the transfer matrix $A(u),B(u),C(u)$
and $D(u)$.
Some of them are listed as
\begin{align}
C(u)B(v)&=g(u,v) \{ A(u)D(v)-A(v)D(u) \}, \\
A(u)B(v)&=f(u,v) B(v) A(u)+g(v,u)B(u)A(v), \label{rel} \\
D(u)B(v)&=f(v,u) B(v) D(u)+g(u,v)B(u)D(v), \label{rel2} \\
[B(u),B(v)]&=[C(u),C(v)]=0. \label{rel3}
\end{align}
The arbitrary $N$-particle state $|\psi(\{u \}) \rangle$
(resp. its dual $\langle \psi(\{u \})|$) 
(not normalized) with $N$ spectral parameters
$\{ u \}=\{ u_1,u_2,\dots,u_N \}$
is constructed by a multiple action
of $B$ (resp. $C$) operator on the vacuum state $|\Omega \rangle$
(respectively, $\langle \Omega|$)
\begin{align}
|\psi(\{u \}) \rangle=\prod_{j=1}^N B(u_j)| \Omega \rangle,\quad
\langle \psi(\{u \})|=\langle \Omega| \prod_{j=1}^N
C(u_j).
\end{align}
Utilizing (\ref{rel}), (\ref{rel2}), (\ref{rel3})
and the action of $A(u)$ and $D(u)$ on the vacuum state
\begin{align}
A(u)|\Omega \rangle=u^M| \Omega \rangle, \ \ \
D(u)|\Omega \rangle=(u-u^{-1})^M | \Omega \rangle,
\end{align}
one can show that the 
state $|\psi(\{u \}) \rangle$
is an eigenstate of the transfer matrix (\ref{transfer})
\begin{align}
\tau(v)| \psi(\{u\}) \rangle
&=\Theta_N(v, \{ u \})| \psi(\{u\}) \rangle, \\
\Theta_N(v, \{ u \})&=
\prod_{j=1}^N \frac{u_j^2}{u_j^2-v^2}
+(1-v^{-2})^M \prod_{j=1}^N \frac{v^2}{v^2-u_j^2},
\end{align}
if the spectral parameters $\{ u \}=\{u_1, u_2, \dots, u_N \}$
satisfy the Bethe ansatz equation
\begin{align}
(1-u_k^{-2})^{-M} u_k^{-2N}=(-1)^{N-1}
\prod_{j=1}^N u_j^{-2},
\label{BAE}
\end{align}
for $k=1,2, \cdots,N$.
One can also show
\begin{align}
\langle \psi(\{u\}) | \tau(v)
=\langle \psi(\{u\}) | 
\Theta_N(v, \{ u \}),
\end{align}
under the constraint (\ref{BAE}).
The eigenvalue of the 
Markov matrix (\ref{markov})
is given by the spectral parameters as
\begin{align}
\mathcal{M}(\{ u \})=\frac{1}{2} \Theta_N^{-1}(1,\{u \})
\frac{\partial}{\partial v} \Theta_N(v, \{u \})|_{v=1}
= \sum_{j=1}^N \frac{1}{u_j^2-1}.
\end{align}
The steady state $| S_N \rangle$ 
($\langle S_N| $) corresponds to
the eigenstate with zero eigenvalue which is given by 
setting the spectral parameters as
$u_1=u_2= \cdots=u_N=\infty$.

The scalar product,
which plays an important role in this paper, 
has the following determinant form \cite{Bo}
\begin{align}
\langle \psi(\{ v \})| \psi(\{ u \}) \rangle
=
\Bigg\{
\prod_{N \ge j>k \ge 1} \frac{v_j v_k}{v_k^2-v_j^2}
\prod_{N \ge l > n \ge 1} \frac{u_l u_n}{u_l^2-u_n^2}
\Bigg\}
\mathrm{det}_N P,
\label{generalscalar}
\end{align}
where $P$ is an $N \times N$ matrix with matrix elements
\begin{align}
P_{jk}=\frac{\Bigg\{
v_j^M(u_k-u_k^{-1})^M \Bigg(
\frac{u_k}{v_j}
\Bigg)^{N-1}
-u_k^M(v_j-v_j^{-1})^M
\Bigg(
\frac{u_k}{v_j}
\Bigg)^{-N+1}
\Bigg\}}
{\frac{u_k}{v_j}-\Big( \frac{u_k}{v_j} \Big)^{-1}}.
\nonumber \\
\label{element}
\end{align}
Taking $ \{ v \}=\{ u \}$, one gets the determinant representation
of the norm 
\begin{align}
\langle \psi(\{u\})|\psi(\{u\}) \rangle 
=\prod_{j=1}^N u_j^{2(N+M-1)}
\prod_{\substack{ l,n=1 \atop l \neq n}}^N \frac{1}{u_l^2-u_n^2}
\mathrm{det}_N Q, \label{norm} 
\end{align}
with $N \times N$ matrix
\begin{align}
Q_{jk}=\frac{N-1+(M-N+1) u_j^{-2}}{1-u_j^{-2}} \delta_{jk}
-(1-\delta_{jk}).
\end{align}
Note that by use of Sylvester's determinant theorem, one can reduce the 
determinant in the above to
\begin{align}
\mathrm{det}_N Q=\prod_{j=1}^N \frac{N+(M-N)u_j^{-2}}{1-u_j^{-2}}
\left(1-\sum_{j=1}^N \frac{1-u_j^{-2}}{N+(M-N)u_j^{-2}}\right).
\end{align}

\section{Form factors and Overlap}
We first formulate the relaxation dynamics of the TASEP to see
what we should evaluate.
Then we evaluate the form factors and the overlap between
the initial state and arbitrary Bethe vector
in the determinant and factorized forms. We consider two
fundamental initial  conditions: the step and alternating initial
conditions.
\subsection{Formulation of the relaxation dynamics}
We formulate the relaxation dynamics of the TASEP
by the algebraic Bethe ansatz.
The time evolution of the expectation 
value for the physical quantity $\mathcal{A}$ starting from an initial 
state $| I_N \rangle$ is defined as
\begin{align}
\langle \mathcal{A} \rangle_t=\langle S_N| \mathcal{A} 
e^{\mathcal{M}t} | I_N \rangle,
\label{expectation}
\end{align}
where $\langle S_N|$ is the left steady state vector
(\ref{leftsteady}).
This definition comes from the fact that the TASEP is a stochastic process,
and the coefficient $P_t(\tau_1,\dots,\tau_M)$ of the state vector
$|\psi(t) \rangle=e^{\mathcal{M}t} | I_N \rangle$
directly gives the probability of being in the state
$|\tau_1,\dots,\tau_M \rangle$.
We decompose the quantity (\ref{expectation})
by inserting the resolution of identity
\begin{align}
I=
\frac{|S_N \rangle \langle S_N|}{Z_N}
+
\sum_{\alpha}
\frac{| \psi_\alpha \rangle \langle \psi_\alpha |}
{
\langle \psi_\alpha| \psi_\alpha \rangle
}. \label{resolution}
\end{align}
Here $\alpha$ labels arbitrary eigenstates except for the steady state.
Then we find the local densities
$\langle n_i \rangle_t=\langle 1-s_i \rangle_t$
and currents $\langle j_i \rangle_t=\langle (1-s_i)s_{i+1} \rangle_t$
are respectively given by
\begin{align}
\langle n_i \rangle_t
&=\frac{N}{M}
+\sum_{\alpha}
\frac{e^{\mathcal{M}_\alpha t} (\langle S_N| \psi_\alpha \rangle-\langle S_N| s_i | 
\psi_\alpha \rangle)\langle \psi_\alpha|I_N \rangle}
{\langle \psi_\alpha|\psi_\alpha \rangle},
 \label{densityexpansion} \\
\langle j_i \rangle_t
&=\frac{N(M-N)}{M(M-1)}  
+\sum_{\alpha}
\frac{e^{\mathcal{M}_\alpha t}(\langle S_N| s_{i+1} | \psi_\alpha \rangle
-\langle S_N| s_i s_{i+1} | \psi_\alpha \rangle)
\langle \psi_\alpha|I_N \rangle} 
{\langle \psi_\alpha|\psi_\alpha \rangle}.
 \label{currentexpansion}
\end{align}
In the same way,
one can also make the same decomposition for
the emptiness formation probability (EFP)
$EFP(i,k)_t=\langle s_is_{i+1} \cdots s_{i+k-1} \rangle_t$,
which gives the probability that the sequence of $k$ sites
($i,(i+1),\dots,(i+k-1)$th sites)
are all unoccupied
\begin{align}
&EFP(i,k)_t \nonumber \\
=&\frac{(M-N)!(M-k)!}{M!(M-k-N)!}
+
\sum_{\alpha}
\frac{e^{\mathcal{M}_\alpha t}
\langle S_N| 
s_is_{i+1} \cdots s_{i+k-1}
| \psi_\alpha \rangle
\langle \psi_\alpha|I_N \rangle} 
{\langle \psi_\alpha|\psi_\alpha \rangle}.
\label{efpexpansion}
\end{align}
From the expressions of the physical quantities
(\ref{densityexpansion}), (\ref{currentexpansion})
and (\ref{efpexpansion}),
one notices that what we should evaluate is the
norm $\langle \psi_\alpha|\psi_\alpha \rangle$,
the form factors $\langle S_N| 
s_is_{i+1} \cdots s_{i+k-1}
| \psi_\alpha \rangle$ and the
overlap between the initial state and
arbitrary Bethe vector $\langle \psi_\alpha|I_N \rangle$. The norm can be obtained as a limit of the scalar product (\ref{generalscalar})
in the determinant form (\ref{norm}).
What remains to be done is the
evaluation of the form factors and the overlap.

\subsection{Form factors}
The form factor for the local operators
$s_i \cdots s_{i+k-1}$ is explicitly given by
the following determinant form
\begin{align}
	&\langle S_N|s_i \cdots s_{i+k-1}|\psi(\{u\}) \rangle
\nonumber \\
=&\prod_{j=1}^N(1-u_j^{-2})^{k+i-1} \prod_{j=1}^N u_j^{M+1}
\prod_{N \ge l > n \ge 1}
\frac{1}{u_l^2-u_n^2} \mathrm{det}_N V^{(M-k)}, \label{formtwo} 
\end{align}
where the $N\times N$ matrix $V$ is written as
\begin{align}
V_{jl}^{(M-k)}=\sum_{n=0}^{j-1} (-1)^n \frac{(M-k)!}{n!(M-k-n)!} 
u_l^{2(j-1-n)}, 
\end{align}
for $1 \le j \le N-1$ and
\begin{align}
V_{Nl}^{(M-k)}=-\sum_{n=N-1}^{M-k} (-1)^n \frac{(M-k)!}{n!(M-k-n)!}
u_l^{-2(n-N+1)}. 
\end{align}
Note that the overlap between the steady state and the
Bethe vector 
$\langle S_N| \psi \rangle$
is obtained by setting
 $i=1, k=0$ in (\ref{formtwo}).
We show (\ref{formtwo}) by applying the approach of \cite{Bo}.
First, let us denote the monodromy matrix
constructed from the $(M-k+1)$-th, $\cdots$, $M$-th sites
and its matrix elements  as
\begin{align}
T_k(u)=\prod_{j=M-k+1}^M L(j|u)
=\left(
\begin{array}{cc}
A_k(u) & B_k(u)  \\
C_k(u) & D_k(u)
\end{array}
\right).
\end{align}
By definition, one has
\begin{align}
&\left(
\begin{array}{cc}
A_{M-k+1}(u) & B_{M-k+1}(u)  \\
C_{M-k+1}(u) & D_{M-k+1}(u)
\end{array}
\right) \nonumber \\
=&
\left(
\begin{array}{cc}
A_{M-k}(u) & B_{M-k}(u)  \\
C_{M-k}(u) & D_{M-k}(u)
\end{array}
\right)
\left(
\begin{array}{cc}
u s_k & \sigma_k^-  \\
\sigma_k^+ & uI-u^{-1}s_k
\end{array}
\right),
\end{align}
from which we get
\begin{align}
B_{M-k+1}(u)&=A_{M-k}(u) \sigma_k^- +B_{M-k}(u)(uI-u^{-1} s_k), \label{eq1} \\
C_{M-k+1}(u)&=C_{M-k}(u) u s_k +D_{M-k}(u) \sigma_k^+. \label{eq2}
\end{align}
Acting both sides of (\ref{eq1}) by $s_k$ from the left,
and (\ref{eq2}) from the right, one has
\begin{align}
s_k B_{M-k+1}(u)&=(u-u^{-1})B_{M-k}(u)s_k \label{eq3}, \\
C_{M-k+1}(u) s_k&=u s_k C_{M-k}(u) \label{eq4}.
\end{align}
Utilizing (\ref{eq3}) and (\ref{eq4}), one can calculate the 
following generalized form factor as
\begin{align}
&\langle \psi(\{v\})|s_1 s_2 \cdots s_k| \psi(\{u\}) 
\rangle \nonumber \\
=&\langle \Omega | \prod_{j=1}^{N-1}C(v_j) C(v_N)
s_1^2 s_2 \cdots s_k B(u_1) \prod_{j=2}^{N}B(u_j)| \Omega \rangle
\nonumber \\
=&\langle \Omega | \prod_{j=1}^{N-1}C(v_j) C(v_N)
s_1 s_2 \cdots s_k s_1 B(u_1) \prod_{j=2}^{N}B(u_j)| \Omega \rangle
\nonumber \\
=&\langle \Omega | \prod_{j=1}^{N-1}C(v_j) v_N s_1 C_{M-1}(v_N)
s_2 \cdots s_k (u_1-u_1^{-1}) B_{M-1}(u_1) s_1
\prod_{j=2}^N B(u_j)| \Omega \rangle
\nonumber \\
=&\prod_{j=1}^N v_j  (u_j-u_j^{-1})
\langle \Omega | \prod_{j=1}^N C_{M-1}(v_j)
s_2 \cdots s_k \prod_{j=1}^N B_{M-1}(u_j)| \Omega \rangle
\nonumber \\
=&\cdots \cdots \nonumber \\
=&\prod_{j=1}^N v_j^k  (u_j-u_j^{-1})^k
\langle \Omega | \prod_{j=1}^N C_{M-k}(v_j)
\prod_{j=1}^N B_{M-k}(u_j)| \Omega \rangle.
\end{align}
Utilizing the cyclic property $s_k=\tau^{k-j} s_j \tau^{j-k}, \tau=\tau(1)$
and the action of $\tau$ on the Bethe vector, one gets
\begin{align}
&\langle \psi(\{v\})|s_i s_{i+1} \cdots s_{i+k-1}| \psi(\{u\}) 
\rangle \nonumber \\
=&\langle \psi(\{v\})|(\tau^{i-1}s_1 \tau^{1-i})
(\tau^{i-1} s_2  \tau^{1-i})
\cdots
(\tau^{i-1} s_k \tau^{1-i})
| \psi(\{u\}) 
\rangle \nonumber \\
=&\langle \psi(\{v\})|\tau^{i-1}s_1 s_2
\cdots s_k \tau^{1-i}
| \psi(\{u\}) 
\rangle \nonumber \\
=&
\Bigg( 
\frac{1-u_j^{-2}}{1-v_j^{-2}}
\Bigg)^{i-1}
\prod_{j=1}^N v_j^k  (u_j-u_j^{-1})^k
\langle \Omega | \prod_{j=1}^N C_{M-k}(v_j)
\prod_{j=1}^N B_{M-k}(u_j)| \Omega \rangle.
\label{formfactorformula}
\end{align}
The form factor
(\ref{formtwo}) can be obtained by taking a limit of
the generalized form factor
(\ref{formfactorformula}).
First, note that the steady state can be obtained as
\begin{align}
|S_N \rangle&=\lim_{\{ u \} \to \infty} \prod_{j=1}^N \tilde{B}_M(u_j)|
\Omega \rangle, \nonumber \\
\langle S_N|&=\lim_{\{ u \} \to \infty} \langle \Omega |
\prod_{j=1}^N \tilde{C}_M(u_j),
\end{align}
where $\tilde{B}_M(u)=u^{-(M-1)} B_M(u)$ and $\tilde{C}_M(u)=u^{-(M-1)}
 C_M(u)$.
We rewrite the generalized form factor
(\ref{formfactorformula}) as
\begin{align}
&\langle \Omega | \prod_{j=1}^N \tilde{C}_M(v_j)
s_i s_{i+1} \cdots s_{i+k-1} \prod_{j=1}^N \tilde{B}_M(u_j)
|\Omega \rangle \nonumber \\
=&
\Bigg( 
\frac{1-u_j^{-2}}{1-v_j^{-2}}
\Bigg)^{i-1}
\prod_{j=1}^N  (1-u_j^{-2})^k
\langle \Omega | 
\prod_{j=1}^N \tilde{C}_{M-k}(v_j)
\prod_{j=1}^N \tilde{B}_{M-k}(u_j)
| \Omega \rangle. \label{beforelimit}
\end{align}
Taking $\{ v \} \to \infty$, one can show \cite{Bo}
\begin{align}
\langle S_N| \prod_{j=1}^N \tilde{B}_{M}(u_j)
| \Omega \rangle=\prod_{k=1}^N u_k^2
\prod_{N \ge l > n \ge 1}
\frac{1}{u_l^2-u_n^2} \mathrm{det}_N V^{(M)},
\label{limitdet}
\end{align}
where
\begin{align}
V_{jl}^{(M)}&=\sum_{n=0}^{j-1} (-1)^n \frac{M!}{n!(M-n)!} 
u_l^{2(j-1-n)},
\ 1 \le j \le N-1, \\
V_{Nl}^{(M)}&=-\sum_{n=N-1}^{M} (-1)^n \frac{M!}{n!(M-n)!}
u_l^{-2(n-N+1)}.
\end{align}
Taking the limit $\{ v \} \to \infty$ in
(\ref{beforelimit}) and inserting
(\ref{limitdet}) to the right hand side,
we have
\begin{align}
&\langle S_N|
s_i s_{i+1} \cdots s_{i+k-1} \prod_{i=1}^N \tilde{B}_M(u_i)
|\Omega \rangle \nonumber \\
=&\prod_{j=1}^N (1-u_j^{-2})^{k+i-1}
\prod_{j=1}^N u_j^2
\prod_{N \ge l > n \ge 1}
\frac{1}{u_l^2-u_n^2} \mathrm{det} V^{(M-k)}.
\end{align}
Changing from $\tilde{B}_M(u)$  to $B_M(u)$, one obtains
the determinant representation for form factors
(\ref{formtwo}).

\subsection{Overlap: Step initial condition}
\begin{figure}[t]
\begin{center}
\includegraphics[width=\textwidth]{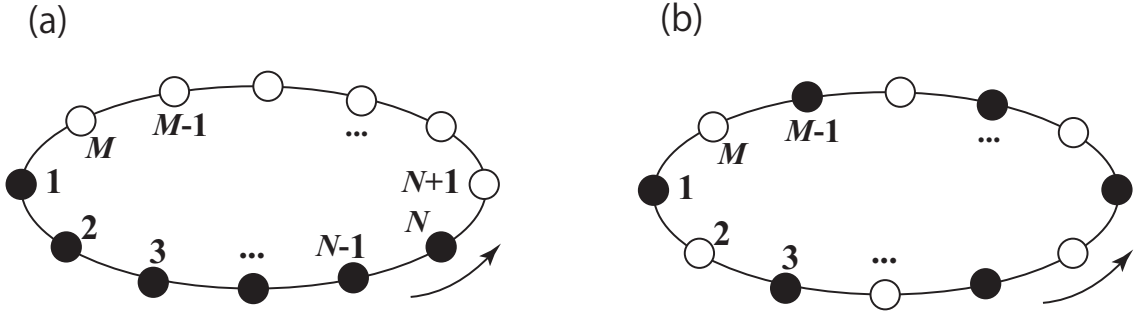}
\end{center}
\caption{(a) The step initial condition and
(b) the alternating initial condition of the TASEP on  a ring.}
\label{initialcondition}
\end{figure}

\begin{figure}[tt]
\begin{center}
\includegraphics[width=0.7\textwidth]{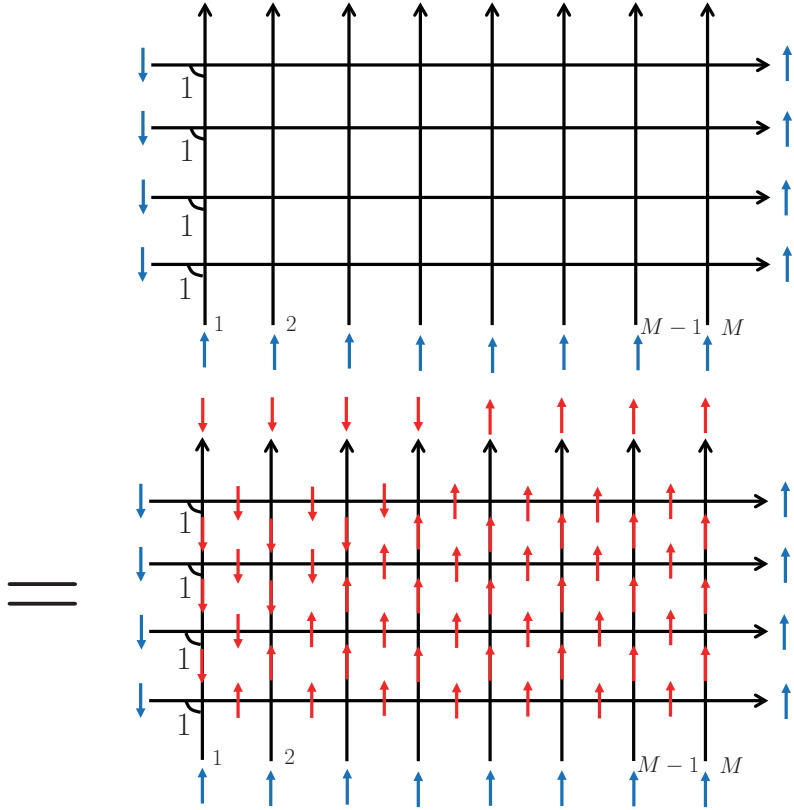}
\end{center}
\caption{Graphical description of 
$|I_N \rangle
=B(1)^N|\Omega \rangle$ for the step
initial condition.
The spins on the bottom line represents the vacuum,
and each row corresponds to the $B$ operator.
Setting the spectral parameter on the bottom line to be one,
one finds the two southwest internal spins freeze since
$\langle \uparrow| \langle \downarrow|_1 
L(1|u=1)|\downarrow \rangle | \uparrow \rangle_1=1$
and
$\langle \downarrow| \langle \uparrow|_1 
L(1|u=1)|\downarrow \rangle | \uparrow \rangle_1=0$.
Repeating the same game
makes one see that all spins freeze.
The spins on the top line is the resultant state of
the action of $B(1)^N$ on the vacuum.
}
\label{initialconstruction}
\end{figure}

What remains to be done is to
evaluate the overlap $\langle \psi_\alpha|I_N \rangle$
between the initial state and arbitrary Bethe vector.
First we consider the step initial condition (see figure
~\ref{initialcondition} (a)) 
 where the half of the system is consecutively occupied by the 
particles and the other half is empty. 
By graphical description of the $L$-operator,
it is easy to see (figure~\ref{initialconstruction})
that the (normalized) initial state
$|I_N\rangle
=|\underbrace{1,\dots,1}_N,\underbrace{0,\dots,0}_{M-N}\rangle$
is given by $|I_N\rangle=B(1)^N|\Omega \rangle$
(see \cite{MC} for the XXZ spin chain).
Note that this initial state is {\em not} the eigenstate
of the Markov matrix.

Then we find the overlap
$\langle \psi(\{v\})|I_N \rangle=
\langle \psi(\{v\})|B(1)^N|\Omega \rangle
=\langle \psi(\{v\})|\psi(\{1\}) \rangle$
can be obtained as a limit $\{u\} \to 1$
of the scalar product formula (\ref{generalscalar}).
One finds
\begin{align}
\langle \psi(\{v\})|I_N \rangle
=&\frac{(-1)^N}{2^{\frac{N(N+1)}{2}}}
\prod_{j=1}^N \frac{(v_j-v_j^{-1})^M}{v_j^{1-2N}}
\prod_{N \ge j> k \ge 1} \frac{1}{v_j^2-v_k^2}
\nonumber \\
&\times \mathrm{det}_N \Bigg(
\frac{1}{(1-v_j)^k}+\frac{1}{(1+v_j)^k}
\Bigg). \label{tochu}
\end{align}
We can furthermore simplify the determinant as
\begin{align}
\mathrm{det}_N \Bigg(
\frac{1}{(1-v_j)^k}+\frac{1}{(1+v_j)^k}
\Bigg) 
=&\mathrm{det}_N \Bigg( 
\frac{2}{(1-v_j^2)^k} \sum_{l=0 \atop l:\mathrm{even}}^k 
\frac{k!}{l!(k-l)!} v_j^l
\Bigg) \nonumber \\
=&\mathrm{det}_N \Bigg(
\frac{2^k}{(1-v_j^2)^k}
\Bigg) \nonumber \\
=&\frac{2^{N(N+1)/2}}{\prod_{j=1}^N (1-v_j^2)^N}
 \mathrm{det}_N ((1-v_j^2)^{N-k}) \nonumber \\
=&\frac{2^{N(N+1)/2} \prod_{N \ge j > k \ge 1 }(v_j^2-v_k^2) }
{\prod_{j=1}^N (1-v_j^2)^N}. \label{simplification}
\end{align}
We made column operation  in the determinant in
the second equality, and used the formula for the Vandermonde determinant
\begin{align}
\mathrm{det}_N(x_j^{N-k})=\prod_{N \ge j > k \ge 1}(x_k-x_j),
\end{align}
in the last equality.
Inserting (\ref{simplification})
into (\ref{tochu}), one gets the following simple form
for the case of the step initial condition
\begin{align}
\langle \psi(\{v\})|I_N \rangle
=\prod_{j=1}^N (v_j-v_j^{-1})^{M-N} v_j^{N-1}. \label{formstep}
\end{align}
Note that this form holds for arbitrary filling.

\subsection{Overlap: Alternating initial condition}
\begin{figure}[tt]
\begin{center}
\includegraphics[width=0.7\textwidth]{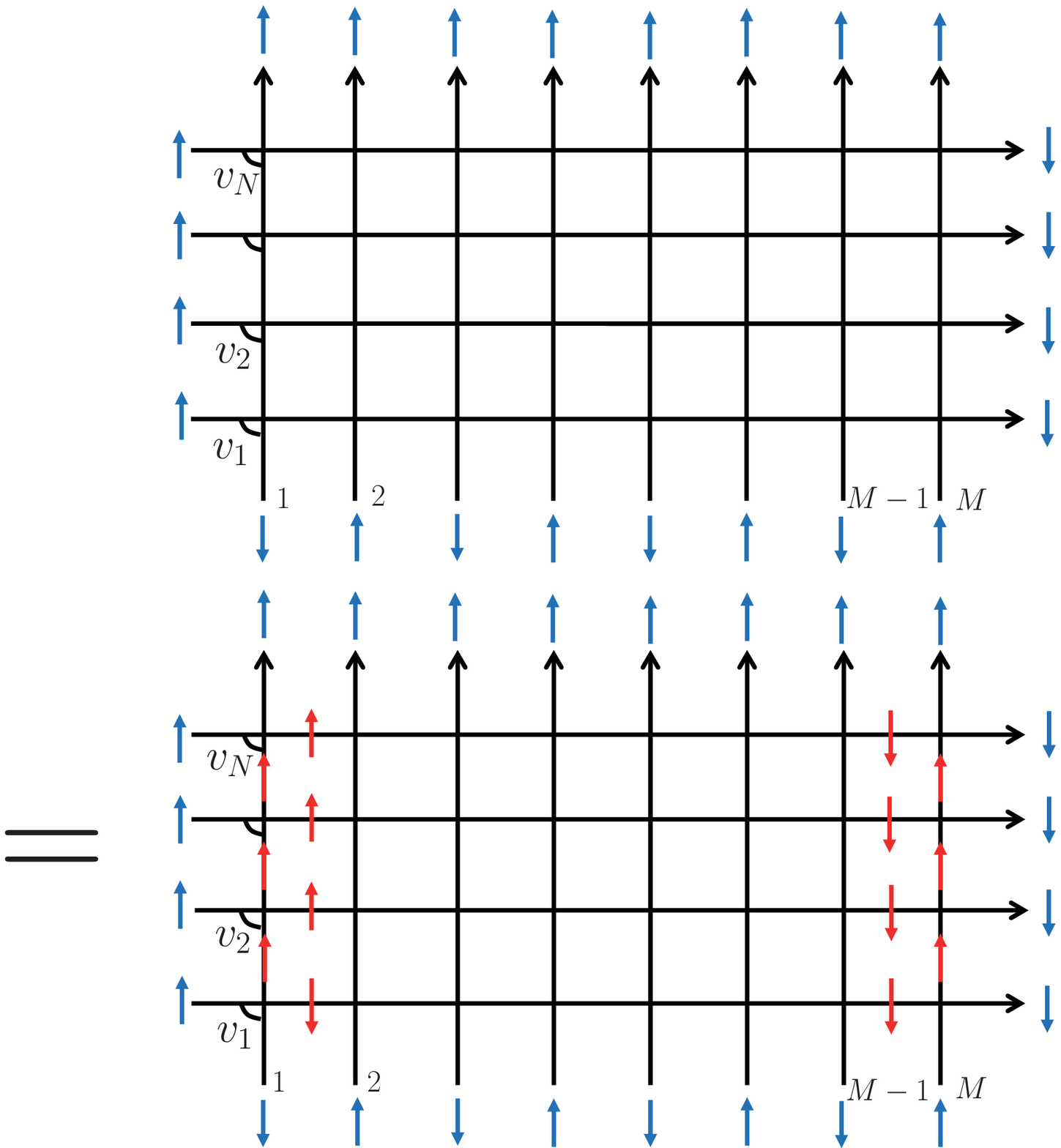}
\end{center}
\caption{Graphical description of
\eqref{alttochu1}
.}
\label{altgraphical1}
\end{figure}

\begin{figure}[tt]
\begin{center}
\includegraphics[width=0.55\textwidth]{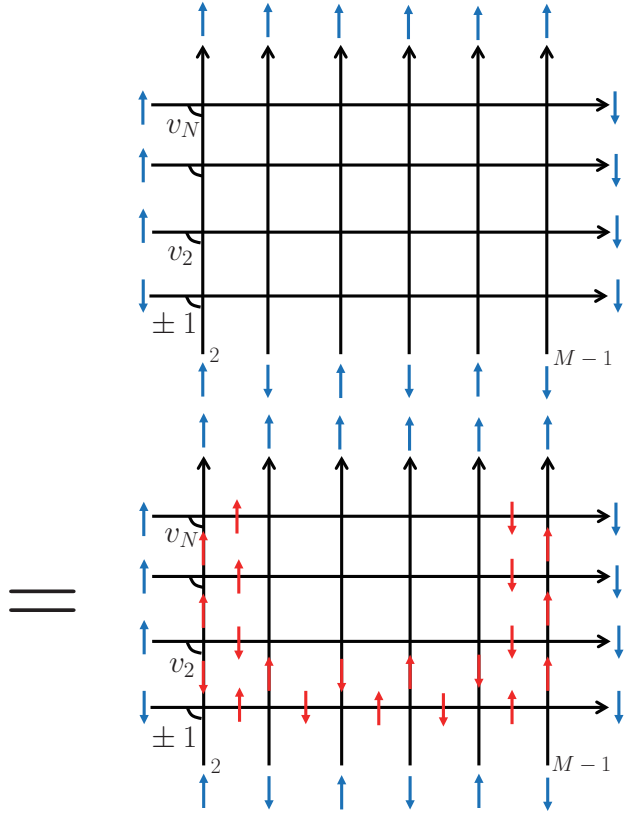}
\end{center}
\caption{Graphical description of
\eqref{rec1}
.}
\label{altgraphical2}
\end{figure}

We now evaluate the overlap for the case of the alternating initial condition
$|I_N \rangle=|1,0,1,0,\dots,1,0 \rangle$
(figure~\ref{initialcondition} (b)),
which
all odd sites are occupied while all even sites are empty.
We consider the half-filling case.
We find the following simple form
\begin{align}
\langle \psi(\{v\})|I_N \rangle
=\prod_{j=1}^N (v_j-v_j^{-1}) 
\prod_{j,k=1 \atop j<k}^N (v_j^2 v_k^2-1). \label{alternatinginitial}
\end{align}
Let us show (\ref{alternatinginitial}).
For convenience, we use the spin notation
(spin up and down states correspond to occupied and empty sites respectively).
First, by representing the left hand side of (\ref{alternatinginitial})
graphically (figure~\ref{altgraphical1}), one finds
\begin{align}
\langle \psi(\{v\})|\downarrow_1 \uparrow_2 \downarrow_3
\uparrow_4 \cdots \downarrow_{M-1} \uparrow_M \rangle
=&\prod_{j=2}^N v_j \prod_{j=1}^N(v_j-v_j^{-1}) 
\mathcal{D}_N(v_1,v_2, \cdots,v_N),
\label{alttochu1}
\\
\mathcal{D}_N(v_1,v_2, \cdots,v_N)
=&\langle \Omega|C(v_N) \cdots C(v_2)D(v_1)|\uparrow_2 \downarrow_3
\uparrow_4 \downarrow_5
\cdots \downarrow_{M-1} \rangle.
\end{align}
We focus on $\mathcal{D}_N(v_1,v_2, \cdots,v_N)$.
Again, by graphical representation (figure~\ref{altgraphical2}),
we note the following recursive relation
\begin{align}
\mathcal{D}_N(\pm1,v_2, \cdots,v_N)=\prod_{j=3}^N v_j
\prod_{j=2}^N(v_j-v_j^{-1}) \mathcal{D}_{N-1}(v_2, \cdots,v_N).
\label{rec1}
\end{align}
Since 
$\langle \psi(\{v\})|\downarrow_1 \uparrow_2 \downarrow_3
\uparrow_4 \cdots \downarrow_{M-1} \uparrow_M \rangle$
is symmetric with respect to $v_1, v_2, \cdots, v_N$ \\
$([B(v_i),B(v_j)]=0)$,
$\mathcal{D}_N(v_1,v_2, \cdots,v_N)$ must be of the form
\begin{align}
\mathcal{D}_N(v_1,v_2, \cdots,v_N)=\prod_{j=2}^N v_j^{-1}
\mathcal{F}_N(v_1,v_2, \cdots,v_N), \label{alttochu2}
\end{align}
where $\mathcal{F}_N(v_1,v_2, \cdots,v_N)$ is a symmetric polynomial
of $v_1, v_2, \cdots, v_N$.
Utilizing (\ref{alttochu2}),
the recursive relation for
$\mathcal{D}_N(v_1,v_2, \cdots,v_N)$ (\ref{rec1})
becomes the one for $\mathcal{F}_N(v_1,v_2, \cdots,v_N)$
\begin{align}
\mathcal{F}_N(\pm1,v_2, \cdots,v_N)=\prod_{j=2}^N(v_j^2-1) 
\mathcal{F}_{N-1}(v_2, \cdots,v_N). \label{rec2}
\end{align}
By symmetry, (\ref{rec2}) extends to
\begin{align}
\mathcal{F}_N(v_1,v_2,\cdots,v_N)|_{v_k=\pm 1}
=\prod_{j=1 \atop j \neq k}^N(v_j^2-1)  \mathcal{F}_{N-1}
(v_1, \cdots, \widehat{v_k} , \cdots, v_N). \label{rec3}
\end{align}
One finds 
\begin{align}
\mathcal{F}_N(v_1,v_2,\cdots,v_N)=
\prod_{j,k=1 \atop j<k}^N (v_j^2 v_k^2-1),
\label{alttochu3}
\end{align}
solves the recursive relation (\ref{rec3}).
Combining (\ref{alttochu1}),(\ref{alttochu2}) and (\ref{alttochu3}),
one gets the factorized polynomial
expression for the overlap (\ref{alternatinginitial})
between the alternating initial state and arbitrary Bethe vector.

\section{Excited states}
In section 2 and 3, we have formulated the dynamics
of the TASEP by the algebraic Bethe ansatz by evaluating
ingredients such as the form factors and overlap.
In this section, we review the
results
of the excitation spectrum of the TASEP on a ring \cite{GM,GM2,deGE1}.
\subsection{Algorithm}
We review the algorithm of computing the
excitation spectrum of the TASEP on a ring.
To this end, we make change of variables
of the spectral parameters from $u_j$ to $Z_j$ 
($u_j^2=(Z_j+1)/(Z_j-1)$)
in this section.
The Bethe ansatz equation can be rewritten as
\begin{align}
(1-Z_k^2)^N=-2^M \prod_{j=1}^N \frac{Z_j-1}{Z_j+1},
\label{betherewritten}
\end{align}
for $k=1,2,\dots,N$.
The eigenvalue of the Markov matrix becomes
\begin{align}
\mathcal{M}(\{Z\})=\sum_{j=1}^N \frac{Z_j-1}{2}.
\end{align}
A simple algorithm was proposed \cite{GM} to calculate
the roots of the Bethe ansatz equation.
Noting the right hand side of (\ref{betherewritten})
does not depend on the index $k$, we define a parameter
$u$ as
\begin{align}
(1-Z_k^2)^N=-\mathrm{e}^{\pi u}.
\end{align}
The roots of this equation are
\begin{align}
Z_m&=-Z_{N+m}=\sqrt{1-y_m}, \label{key} \\
y_m&=\mathrm{e}^{2 \pi (u+i)/M}\mathrm{e}^{4 \pi i(m-1)/M},
\end{align}
for $m=1,2,\cdots,N$,
and $\{y\}=\{y_1,y_2,\dots,y_N \}$ satisfy
\begin{align}
0 \le \mathrm{arg}(y_1)<\mathrm{arg}(y_2)<
\cdots \mathrm{arg}(y_N)<2 \pi.
\end{align}
It was proposed in \cite{GM} that choosing a
sequence of quantum numbers $\{c(1),c(2),\dots,c(N) \}$
satisfying $1 \le c(1) < c(2) \cdots < c(N) \le M$
and determining the parameter $u$ from
\begin{align}
\mathrm{e}^{\pi u}=
2^M \prod_{j=1}^N \frac{Z_{c(j)}-1}{Z_{c(j)}+1},
\end{align}
self-consistently
by numerical iteration,
one gets the Bethe roots 
$\{ Z_{c(1)},Z_{c(2)},\dots,Z_{c(N)} \}$.

\subsection{Excited states}
By numerical calculation,
we observe the following
types of Bethe roots
are the three lowest excited states.
We mean a lower-lying excited state
by a state whose real part of its
corresponding eigenvalue
of the Markov matrix is closer to zero.
Here, we impose the ansatz
that the eigenvalues of the low-lying excited states behave as
$\mathrm{ln} (\mathrm{Re} \mathcal{M})
=\alpha-\beta \mathrm{ln} M, \beta=3/2$ by the following reasons.
Since the lowest excited states believed to be true
behave in this way, the exponents $\beta$ for the other excited states
should not be $\beta>3/2$ for $M$ large enough
since no crossing across the lowest excited states is allowed.
Next, by estimating several low-lying excited states
by numerical observations for $M \sim 20$ and
making finite size scaling analysis of them for large $M$,
we find the exponents for all of those states are close to $\beta=3/2$,
which is the reason why we impose the ansatz.
\\
$(1)$Type I  \cite{GM,GM2,deGE1} \\
The Bethe roots corresponding to the quantum numbers
\begin{align}
&\{ c(j)=j \ \mathrm{for} \ j=1,\cdots,N-1, \ c(N)=N+1 \}, \\
&\{ c(j)=j+1 \ \mathrm{for} \ j=1,\cdots,N-1, \ c(N)=2N \},
\end{align}
give the lowest excited states. 
The simplest fitting from $M=256,512,1024$ gives
$\mathrm{ln}(-\mathcal{M}_{\mathrm{1st}})=1.89793-1.50351 \mathrm{ln}M$,
which implies the KPZ scaling, i.e. $\mathcal{M}=C M^{-3/2}$ for 
$M \gg 1$.
\\
$(2)$Type II \\
The second lowest excited state is given by the 
quantum number
\begin{align}
\{ c(j)=j+1 \ \mathrm{for} \ j=1,\cdots,N-2, \ c(N-1)=N+1, \ c(N)=2N \},
\end{align}
for $M$ large enough.
Finite size scaling from $M=256,512,1024$ shows
$\mathrm{ln}(-\mathcal{M}_{\mathrm{2nd}})=2.81592-1.505768\mathrm{ln}M$,
which shows the KPZ scaling again.
This state will be important for the case of 
the alternating initial condition.
\\
$(3)$Type III \\
The Bethe roots associated with the following four
quantum numbers
\begin{align}
&\{ c(j)=j \ \mathrm{for} \ j=1,\cdots,N-1, \ c(N)=N+2 \}, \\
&\{ c(j)=j \ \mathrm{for} \ j=1,\cdots,N-2,  \ c(N-1)=N, \ c(N)=N+1 \},
\\
&\{ c(1)=1, \ c(j)=j+1 \ \mathrm{for} \ j=2,\cdots,N-1, \ c(N)=2N \},
\\
&\{ c(j)=j+1 \ \mathrm{for} \ j=1,\cdots,N-1, \ c(N)=2N-1 \},
\end{align}
give the third lowest excited states. 
Conducting the finite size scaling from $M=256,512,1024$
shows $\mathrm{ln}(-\mathrm{Re}\mathcal{M}_{\mathrm{3rd}})
=2.86964-1.50347 \mathrm{ln} M$, which also belongs to the
KPZ scaling.

\section{Step initial condition}
In this and the next section, we examine the
relaxation times by studying long time asymptotics of the
local current and emptiness formation probability.
We consider the step initial condition in this section. \\
We first analyze the local current.
In the long time, the local current behaves as
\begin{align}
\langle j_k \rangle_t \rightarrow&
\frac{N(M-N)}{M(M-1)}
+
A_{\mathrm{1st}}(j_k) \mathrm{e}^{\mathcal{M}_{\mathrm{1st}}t}
\nonumber \\
&+
A_{\mathrm{2nd}}(j_k) \mathrm{e}^{\mathcal{M}_{\mathrm{2nd}}t}
+
A_{\mathrm{3rd}}(j_k)
\mathrm{e}^{\mathrm{Re}\mathcal{M}_{\mathrm{3rd}}t}
\mathrm{cos}
(\mathrm{Im}\mathcal{M}_{\mathrm{3rd}}t+\delta_k)
\ \mathrm{as} \ t \rightarrow \infty.
\end{align}
The simplest fitting from
$M=256,512,1024$ shows
$\mathrm{ln}[A_{\mathrm{1st}}(j_N)]
=0.854409-0.9968579\mathrm{ln}M$,
$\mathrm{ln}[-A_{\mathrm{2nd}}(j_N)]
=1.1875099-0.9838921\mathrm{ln}M$ 
and
$\mathrm{ln}[|A_{\mathrm{3rd}}(j_N)|]=2.20350-0.98345 \mathrm{ln}M$
from which one concludes
$A_{\mathrm{1st}}(j_N) \propto M^{-1}$,
$A_{\mathrm{2nd}}(j_N) \propto M^{-1}$
and
$|A_{\mathrm{3rd}}(j_N)| \propto M^{-1}$.
Especially, $A_{\mathrm{1st}}(j_N) \propto M^{-1}$
means that $A_{\mathrm{1st}}(j_N) \neq 0$
as long as the total number of sites $M$ is finite,
confirming the relaxation time of the step initial condition
$\tau_{\mathrm{step}}$ is determined by the
lowest eigenvalue of the Markov matrix which shows the KPZ scaling
$\tau_{\mathrm{step}}=-\mathrm{Re}(1/\mathcal{M}_{\mathrm{1st}})
\propto M^{3/2}$.

We can also examine the asymptotic amplitudes of the EFP.
\begin{align}
EFP(i,k)_t \rightarrow \frac{(M-N)!(M-k)!}{M!(M-k-N)!}
+A(EFP(i,k)) \mathrm{e}^{\mathcal{M}_{\mathrm{1st}}t}
\ \mathrm{as} \ t \rightarrow \infty.
\end{align}
Table~\ref{math-tab2} is the results of the finite size scaling
of the asymptotic amplitudes.
One observes $A(EFP(i,k)) \propto M^{-\alpha_k}$, and $\alpha_k$
becomes smaller as the length $k$ becomes longer.

\begin{table}
\caption{\label{math-tab2}Table of the asymptotic amplitudes
vs total number of sites (step initial condition).}
\begin{tabular*}{\textwidth}
{l|l}
\hline
$(i,k)$ & $\mathrm{ln}[-A(EFP(i,k))]$  \\ 
\hline
$(N,1)$ & $1.93798-1.49332 \mathrm{ln}M$  \\ 
$(N-1,2)$ & $1.35223-1.05592 \mathrm{ln}M$  \\ 
$(N-2,3)$ & $1.42851-1.01683 \mathrm{ln}M$ \\ 
$(N-3,4)$ & $1.22155-0.99148 \mathrm{ln}M$ \\ 
$(N-4,5)$ & $0.84935-0.96772 \mathrm{ln}M$ \\ 
$(N-5,6)$ & $0.36629-0.94290 \mathrm{ln}M$ \\ 
$(N-6,7)$ & $-0.19885-0.91613 \mathrm{ln}M$ \\ 
$(N-7,8)$ & $-0.82903-0.88705 \mathrm{ln}M$ \\ 
$(N-8,9)$ & $-1.51332-0.85549 \mathrm{ln}M$ \\ 
$(N-9,10)$ & $-2.24430-0.82136 \mathrm{ln}M$ \\ 
\hline
\end{tabular*}
\end{table}

\section{Alternating initial condition}
Next we study the alternating initial condition.
Surprisingly, 
we find that the asymptotic amplitudes of the local currents
associated with the lowest (Type I) and the third lowest
(Type III) excited states vanish.
This can be shown as follows.
Rewriting the overlap (\ref{alternatinginitial})
in terms of the spectral parameters $Z_{c(j)}$ in section 4,
the following terms
\begin{align}
\prod_{j,k=1 \atop j<k}^N(Z_{c(j)}+Z_{c(k)}),
\end{align}
appear.
Let us take a look at one of the excited states of Type I
$\{ c(j)=j \ \mathrm{for} \ j=1,\cdots,N-1, \ c(N)=N+1 \}$,
for example.
There exists a term  $Z_1+Z_{N+1}$ since $c(1)=1,c(N)=N+1$.
However, this term is equal to zero
since it is one of the cases of
$Z_k+Z_{N+k}=0$ (\ref{key}).
Not only the excited states of Type I
but also for a large number of low-lying excited states,
Type III, for example, have terms $Z_k+Z_{N+k}$ which 
eventually become zero.
These states do not
make any contributions to the relaxation dynamics.
Instead, we find that the second lowest excited state
(Type II) determines the relaxation time $\tau_{\mathrm{alt}}$
for the case of the alternating initial condition.
Note that the eigenvalue corresponding to
the second lowest excited state
obeys the KPZ scaling
$\tau_{\mathrm{alt}}=-\mathrm{Re}(1/\mathcal{M}_{\mathrm{2nd}})
\propto M^{3/2}$ (see Section 4.2).
Thus the local current asymptotically behaves as
\begin{align}
\langle j_k \rangle_t& \rightarrow
\frac{N(M-N)}{M(M-1)}
+
B_{\mathrm{2nd}}(j_k) \mathrm{e}^{\mathcal{M}_{\mathrm{2nd}}t} \ \mathrm{as} \
t \to \infty.
\end{align}
The fitting from
$M=256,512,1024$ shows
$\mathrm{ln}[B_{\mathrm{2nd}}(j_k)]=0.23825-0.99819\mathrm{ln}M$
for both $k$ odd (initially occupied) and even (initially empty).
From this, one concludes $B_{\mathrm{2nd}}(j_k) \propto M^{-1}$,
which is the same with the step initial condition.

Next, we analyze the EFP.
Again, we find the lowest and third lowest excited states do not
contribute.
\begin{align}
EFP(i,k)_t \rightarrow \frac{(M-N)!(M-k)!}{M!(M-k-N)!}
+B(EFP(i,k)) \mathrm{e}^{\mathcal{M}_{\mathrm{2nd}}t}
\ \mathrm{as} \ t \rightarrow \infty.
\end{align}

\begin{table}
\caption{\label{math-tab2}Table of the asymptotic amplitudes
vs total number of sites (alternating initial condition).}
\begin{tabular*}{\textwidth}
{l|l}
\hline
$(i,k)$ & $\mathrm{ln}[-B(EFP(i,k))]$  \\ 
\hline
$(1,2)$ & $0.23825-0.99819 \mathrm{ln}M$  \\ 
$(1,3)$ & $0.63787-0.99736 \mathrm{ln}M$ \\ 
$(1,4)$ & $0.60321-0.99248 \mathrm{ln}M$ \\ 
$(1,5)$ & $0.35850-0.98402 \mathrm{ln}M$ \\ 
$(1,6)$ & $-0.01838-0.97179 \mathrm{ln}M$ \\ 
$(1,7)$ & $-0.49028-0.95602 \mathrm{ln}M$ \\ 
$(1,8)$ & $-1.03631-0.93678 \mathrm{ln}M$ \\ 
$(1,9)$ & $-1.64345-0.91417 \mathrm{ln}M$ \\ 
$(1,10)$ & $-2.30316-0.88825 \mathrm{ln}M$ \\ 
\hline
\end{tabular*}
\end{table}
One observes $B(EFP(i,k)) \propto M^{-\beta_k}$, and $\beta_k$
becomes smaller as the length $k$ becomes longer.
This behavior is similar with the step initial condition.

\section{Conclusion}
In this paper, we studied the long time asymptotics of the
relaxation dynamics of the totally asymmetric simple exclusion process.
We examined the local currents by the algebraic Bethe ansatz method.
By evaluating the asymptotic amplitudes, we find the relaxation times
starting from the step and alternating initial conditions are
governed by different eigenvalues of the Markov matrix.
The relaxation time of the step initial condition $\tau_{\mathrm{step}}$
is given by the nonzero eigenvalue of the Markov matrix
with the first largest real part $\mathcal{M}_{\mathrm{1st}}$ as
$\tau_{\mathrm{step}}=-\mathrm{Re}(1/\mathcal{M}_{\mathrm{1st}})$.
On the other hand, the relaxation time of the alternating initial condition
$\tau_{\mathrm{alt}}$ is given by the nonzero eigenvalue of the Markov matrix
with the second largest real part $\mathcal{M}_{\mathrm{2nd}}$ as
$\tau_{\mathrm{alt}}=-\mathrm{Re}(1/\mathcal{M}_{\mathrm{2nd}})$
for large number of total sites.
The difference between the step and alternating initial conditions
is observed in another context: the current fluctuation.
The GUE Tracy-Widom distribution appears for the case of the step
initial condition \cite{Jo}.
On the other hand, the current fluctuation for the alternating initial condition
is described by the GOE Tracy-Widom distribution \cite{BR,PS}.
Our results of the relaxation times is another aspect of
the difference between the step and alternating initial conditions.

In this paper, we examined the dynamics of the TASEP by use of the
algebraic Bethe ansatz method.
It is interesting to study other correlation functions and
extend to other cases like open boundary, for example.
The recent advances \cite{CRS,CRS2} might be helpful for this case.
It would also be valuable to study the link with the random matrix theory.
The case examined in this paper is when the time is large enough,
while the case obtained by the random matrix theory is
when the time and the size of the system are comparable.
These two results are considered to be supplementary to each other,
and unifying the results by the Bethe ansatz would be an interesting
problem.

\section*{Acknowledgment}
We thank T. Imamura for useful discussions and comments.
The present work was partially supported
by Grants-in-Aid  for Scientific Research (C) No. 24540393 and 
JSPS Fellows from Japan Society for the Promotion of Science.

\end{document}